# Extending the north-east limit of the chart of nuclides.

J. Benlliure[1,], H. Alvarez-Pol[1], T. Kurtukian-Nieto[1,a], K.-H. Schmidt[2], L. Audouin[3], B. Blank[4], F. Becker[2], E. Casarejos[1], D. Cortina-Gil[1], T. Enqvist[2,b], B. Fernández[5], M. Fernández-Ordóñez[1,c], J. Giovinazzo[4], D. Henzlova[2,d], A.R. Junghans[6], B. Jurado[4], P. Napolitani[3], J. Pereira[1,d], F. Rejmund[5], O. Yordanov[2,e]

*[1] Universidade de Santiago de Compostela, E-15782 Santiago de Compostela, Spain, [2]Gesellschaft fuer Schwerionenforschung, D-64291Darmstadt, Germany, [3]Institut de Physique Nucleaire, F-91406 Orsay Cedex, France, [4]Centre d'Etudes Nucleaires Bordeaux-Gradignan, F33175 Gradignan Cedex, France, [5]GANIL BP55025,F-14076 Caen, France*

[a]Present address: CENBG, Le Haut-Vigneau, BP 120, F-33175 Gradignan Cedex, France, [b]Present address: CUPP project, P.O. Box 22, FIN-8680, Pyhäsalmi, Finland, [c]Present address: CIEMAT, Madrid, Spain, [d]Present address: NSCL, East Lansing, Michigan 48824-1321, USA, [e]Present address: INRNE, Sofia, Bulgaria

**The existence of nuclei with exotic combinations of protons and neutrons provides fundamental information on the forces acting between nucleons. The maximum number of neutrons a given number of protons can bind, neutron drip line[1], is only known for the lightest chemical elements, up to oxygen. For heavier elements, the larger its atomic number, the farther from this limit is the most neutron-rich known isotope. The properties of heavy neutron-rich nuclei also have a direct impact on understanding the observed abundances of chemical elements heavier than iron in our Universe. Above half of the abundances of these elements are thought to be produced in rapid-neutron capture reactions, r-process, taking place in violent stellar scenarios[2] where heavy neutron-rich nuclei, far beyond the ones known up today, are produced. Here we present a major step forward in the**

**production of heavy neutron-rich nuclei: the discovery of 73 new neutron-rich isotopes of chemical elements between tantalum (Z=72) and actinium (Z=89). This result proves that cold-fragmentation reactions[3] at relativistic energies are governed by large fluctuations in isospin and energy dissipation making possible the massive production of heavy neutron-rich nuclei, paving then the way for the full understanding of the origin of the heavier elements in our Universe. It is expected that further studies providing ground and structural properties of the nuclei presented here will reveal further details on the nuclear shell evolution along Z=82 and N=126, but also on the understanding of the stellar nucleosyntheis r-process around the waiting point at A~190 defining the speed of the matter flow towards heavier fissioning nuclei.**

The access to unexplored regions of the chart of nuclides far from stability offers unique opportunities in modern nuclear physics but also in nuclear astrophysics. Studying ground and structural properties of nuclei close to the drip lines, crossing shell gaps or in particular regions of deformation represents the most important testing ground for improving the description of the nuclear force acting in the nuclear many body system. Some examples of new phenomena constraining the nuclear interaction, that appear far from stability, are new shell gaps[4], evidencing the role of the tensor term in the nuclear force[5], or abnormal matter distributions such as haloes or neutron skins[6].

Many of the stellar nucleosynthesis processes also involve nuclei far from stability, in particular, those taking place in violent scenarios like novae or super-novae, leading to the production of the heaviest nuclei in the Universe[7]. Clear examples are the rapid-proton capture, rp-process, or the rapid neutron capture, r-process, where nuclei approaching the proton and neutron drip lines, respectively, are produced. The full understanding of these nucleosynthesis processes requires then, experimental information on the bulk and structural properties of many of those nuclei.

The number of nuclear species bound by the nuclear force can be estimated using theoretical descriptions of the nuclear binding energy. According to the finite range droplet model[8], around 6000 nuclear systems could be bound by the nuclear force. The last edition of the chart of nuclides[9] includes up to 2962 nuclei observed so far. The known nuclei reach the proton drip line up to mercury (Z=80), while the exact location of the neutron drip line is only known up to oxygen (Z=8). Then, the heavier the element the further is the last known neutron-rich isotope from the estimated position of the drip line. This situation becomes dramatic for elements above dysprosium (Z=66), where the present limit of the chart of nuclides is just few neutrons away from stability. The presence in this area of the chart of nuclides of regions of deformation, the last known nuclear shells (Z=82, N=126) and the last waiting point of the stellar nucleosynthesis r-process, provides strong motivation for the present efforts for extending the north-east limit of the nuclear chart.

The production of nuclei far from stability requires two conditions: an optimum reaction mechanism populating the nuclear species of interest with the largest possible rate and an efficient experimental technique to identify and separate those nuclei. The choice of the reaction mechanism is limited by the possible ions that can be accelerated, their intensities and the maximum useful target thickness according to the projectile energy. A large fraction of the proton drip line is reachable using spallation, fragmentation and fusion-evaporation reactions[10]. Light nuclei at or close to the neutron drip line are accessible by fragmentation and spallation reactions[11,12]. Fission is successfully used for producing medium-mass neutron-rich nuclei[13], although the fragmentation of $^{136}$Xe is also competitive in some particular cases[14], and the combination of both reaction mechanisms has been proposed as a future option[15]. Some years ago it was proposed to use fragmentation reactions at high energy to overcome the present limitation for producing heavy neutron-rich nuclei[16]. The main argument was that this process is governed by large fluctuations in the neutron excess of the final reaction products but at the same time profit from relatively large useful target thicknesses. The present work aimed at the verification of these arguments.

From a technical point of view two different approaches are proposed for the production of nuclei far from stability, the *in-flight* fragmentation/fission of high energy projectiles and the *isotopic separation on-line* (ISOL) of residual nuclei produced in spallation or fission reactions induced by low/moderate energy particles. The *in-flight* method takes advantage of the inverse kinematics that makes very efficient the magnetic identification and selection of the nuclear species of interest. The *ISOL* method profits from higher primary beam intensities and thicker targets, although the separation method is chemically dependent and the breading and post-acceleration processes suffers from reduced efficiencies. Despite recent efforts[17,18], the *ISOL* technology is limited for the production of heavy neutron-rich nuclei because of the refractory nature of some of them, between Hf and Pt, while heavier ones are contaminated by strongly produced isobars with higher Z. However, it is expected that *in-flight* fragmentation of heavy stable beams gives us access to this region of the chart of nuclides provided that sufficient beam intensities are available.

The experiment we present here aimed at investigating the production of heavy neutron-rich nuclei taking advantage of the relativistic heavy-ion beams produced with the SIS synchrotron at GSI, Darmstadt. Beams of $^{238}$U and $^{208}$Pb ions were accelerated up to 1 A GeV with a typical intensity of $10^8$ ions/s, that was continuously monitored using a secondary electron monitor. The beams extracted from the synchrotron were then directed to a beryllium production target with a thickness of 1600 mg/cm$^2$. Projectile fragments, flying forward, were then separated and unambiguously identified using the magnetic spectrometer FRagment Separator (FRS)[19]. The FRS is a zero-degree, magnetic spectrometer with a resolving power $B\rho/\Delta B\rho \sim 1500$, a momentum acceptance $\Delta p/p \sim 1.5\%$ and an angular acceptance of 15 mrad for the central trajectories. The symmetric two-stages of the spectrometer, depicted in figure 1, guarantee a final achromatic image plane. The first stage of the spectrometer provides a separation according to the mass number over ionic charge ratio of the transmitted fragments. The use of a properly shaped aluminium degrader at the intermediate image plane, preserving the achromatic character of the spectrometer, provides an additional selection in the second stage of the spectrometer according to the atomic number of the transmitted fragments[20]. Therefore, the appropriate tuning of the magnets in both sections of the spectrometer allows us to select a given region of the chart of nuclides.

Several detectors placed along the spectrometer make possible the identification of the transmitted nuclei. Two plastic scintillators, located at the intermediate and final image planes, equipped with two photomultipliers at both ends of each scintillator provide the dispersive coordinate of the trajectories at both image planes but also their time of flight. These two positions allow us to determine the magnetic rigidity of the fragments while the time-of-flight represents a measurement of their velocity. Two additional multi-sampling ionisation chambers (MUSIC) placed at the final image plane are used to determine the atomic number of the nuclei.

A particular challenge in the identification of heavy neutron-rich nuclei is the separation of nuclei produced with different atomic charge states. Indeed, any neutron-rich nuclei (A,Z) could be heavily contaminated by a minor fraction of (A-3,Z) hydrogen-like ions, having almost the same magnetic rigidity but a much larger production cross section. Two conditions seem then mandatory for the unambiguous identification of heavy neutron-rich nuclei: minimize the fraction of none fully stripped ions and force and identify changes in the atomic charge state of the transmitted nuclei.

In the present experiment both requirements were fulfil by using beams at relativistic energies but also niobium stripper foils at the target, at the intermediate image plane and in between the two MUSIC chambers. By measuring the difference in magnetic rigidity induced by the atomic charge changing at the intermediate image plane of the spectrometer and the different energy losses in both MUSIC chambers, also due to atomic charge changes at the stripper between both chambers, we could guarantee a ratio of contaminants below 30%.

Using this method, and scanning with the FRS magnets the region of magnetic rigidities larger than 12.91 Tm, we were able to produce and identify more than 130 heavy neutron-rich isotopes of elements between tantalum (Z=72) and actinium (Z=89), 73 of them for the first time, using roughly five measurement days per beam. In figure 2 we show a single identification matrix produced combining the more than 20 different magnetic tunings of the FRS with both beams where the present limits of the chart of nuclides are indicated by solid lines. The condition used to validate the identification of a new isotope was to observe at least five events with the corresponding value of atomic and mass number.

We could also determine the production cross sections of all identified nuclei normalising the measured yields to the beam intensity and target thickness, and applying corrections due to the transmission through the spectrometer, detection efficiency, secondary reactions in all layers of matter along the FRS and the data acquisition dead time[21]. The observed nuclei are depicted in figure 3 on top of a chart of nuclides where the colour code indicates the production cross sections.

The present data show how fragmentation reactions at relativistic energies produce nuclei far from stability in a wide range of chemical elements covering the gaps between the possible nuclear species that can be used as projectiles in this region of the chart of nuclides, $^{238}$U, $^{208}$Pb and $^{186}$W. Moreover, nuclei along the projectile's isotonic lines are significantly populated. These are nuclei produced in cold-fragmentation reactions where only protons are abraded from the projectile, up to six in this case, while the resulting pre-fragments are excited below the particle evaporation threshold. This particular reaction channel demonstrates the large fluctuations governing the fragmentation process both in isospin but also in excitation energy, being the key issue in the production of the most neutron-rich nuclei.

The dotted and thick-solid lines in figure 3, showing the most neutron-rich nuclei known in 1958 and before the present experiments, clearly indicate how the use of cold-fragmentation reactions at relativistic energies represents a major step forward in the production of heavy neutron-rich nuclei since this reaction mechanism made possible to produce and identify more nuclei in this region of the chart of nuclides that during the past 50 years of research. Indeed, after these measurements[22,23,24], several experiments could already investigate ground and structural properties of some of these nuclei[25,26,27], taking advantage of the cold-fragmentation mechanism for their production.

In summary, our work shows that projectile-fragmentation reactions at relativistic energies investigated in-flight constitute an optimum tool for exploring the north-east region of the chart of nuclides. In the present case we could synthesize for the first time 73 new heavy neutron-rich nuclei opening a completely unexplored territory to the new radioactive beam facilities under construction in Europe (FAIR), in the United States (FRIB) or in Japan (RIBF).

1. Upper and lower limits in the number of neutrons for bound nuclei.

2. Above half of the elements heavier than iron are thought to be produced in violent stellar environments, like supernovae type II or binary neutron mergers, where large neutron fluxes ($10^{24-26}$ cm$^{-3}$) ignite the rapid capture of neutrons. The competition between neutron capture, photo-disintegration and β⁻-decays leadd to the production of heavy nuclei.

3. These are peripheral reactions between heavy ions at relativistic energies where the non-overlapping region between projectile at target leads to projectile/target remanants with large excitation energy and moderate angular momentum. The large fluctuations in excitation energy and isospin governing fragmentation reactions made possible to populate cold-fragmentation channels where mostly protons are abraded from the projectile while the final remanant gains a value of excitation energy below the particle evaporation threshold. This reaction channels leads to the production of the most neutron-rich nuclei accesible to fragmentation reactions.

4. Sorlin, O. & Porquet, M.-G. Nuclear magic numbers: new features far from stability. *Prog. Part. Nucl. Phys.* **61,** 602–673 (2008).

5. Otsuka, T., Matsuo, T. & Abe, D. Mean field with tensor force and shell structure of exotic nuclei. *Phys.Rev. Lett.* **97,** 162501 (2006).

6. Hansen, P.G. Nuclear structure at the drip lines. *Phys. Scr. T* **32,** 21 (1990).

7. Grawe, H., Langanke & K., Martinez-Pinedo, G. Nuclear structure and astrophysics. *Rep. Prog. Phys.* **70**, 1525 (2007).

8. Moller, P., Nix, J.R., Myers, W.D. &Swiatecki, W.J. Nuclear ground-state masses and deformations. *At. Data Nucl. Data Tables* **59**, 185-381 (1995).

9. Magill, J., Pfennig, G. & Galy, J. Chart of the nuclides, 7$^{th}$ edition (2006).

10. Adamian, G.G., Antonenko, N.V., Scheid, W. & Zubov, A.S. Possibilities of production of neutron-deficient isotopes of U, Np, Pu, Am, Cm, and Cf in complete fusion reactions. *Phys. Rev. C* **78,** 044605 (2008).

11. Mocko, M. *et al*. Projectile fragmentation of $^{40}$Ca, $^{48}$Ca, $^{58}$Ni, and $^{64}$Ni at 140 MeV/nucleon. *Phys. Rev. C* **74**, 054612 (2006).

12. Ravn, H.L. Radioactive ion beams available at on-line mass separators. *Nucl. Instrum. Methods B* **26**, 72 (1987).

13. Bernás, M. *et al*. Projectile fission at relativistic velocities: a novel and powerful source of neutron-rich isotopes well suited for in-flight isotopic separation. *Phys. Lett. B* **331**, 19 (1994).

14. Benlliure, J. *et al*. Production of medium-mass neutron-rich nuclei in reactions induced by $^{136}$Xe projectiles at 1 A GeV on a beryllium target. *Phys. Rev. C* **78**, 054605 (2008).

15. Helariutta, K., Benlliure, J., Ricciardi, M.V., & Schmidt, K.-H. Model calculations of a two-step reaction scheme for the production of neutron-rich secondary beams. *Eur. Phys. J. A* **17**, 181 (2003).

16. Benlliure, J. *et al.* Production of neutron-rich isotopes by cold fragmentation in the reaction 197Au + Be at 950 A MeV. *Nucl. Phys. A* **660**, 87 (1999).

17. De Witte, H. *et al.* First observation of the β decay of neutron-rich $^{218}$Bi by the pulsed-release technique and resonant laser ionization. *Phys. Rev. C* **69**, 044305 (2004).

18. Neidherr, D. *et al.* Discovery of $^{229}$Rn and the structure of the heaviest Rn and Ra isotopes from penning-trap mass measurements. *Phys. Rev. Lett.* **102,** 112501 (2009).

19. Geissel, H. *et al.* The GSI projectile fragment separator (FRS): a versatile magnetic system for relativistic heavy ions. *Nucl. Instrum. and Methods B* **70**, 286 (1992).

20. Schmidt, K.-H. *et al.* The momentum-loss achromat — A new method for the isotopical separation of relativistic heavy ions. *Nucl. Instrum. and Methods* A, **260**, 287 (1987).

21. Casarejos, E. *et al.* Isotopic production cross sections of spallation-evaporation residues from reactions of $^{238}$U(1A GeV) with deuterium. *Phys. Rev. C* **74**, 044612 (2006).

22 Kurtukian-Nieto, T. Production and β-decay half-lives of heavy neutron-rich nuclei approaching the stellar nucleosynthesis r-process path around A~195. PhD University of Santiago de Compostela (2007).

23. Benlliure, J. *et al.* Production of Heavy neutron-rich nuclei south of lead. *Eur. Phys. J. Special Topics* **150**, 309 (2007).

24. Alvarez-Pol, H. *et al.* Production cross sections of neutron-rich Pb and Bi isotopes in fragmentation of $^{238}$U. *Eur. Phys. J. A* **42**, 485 (2009).

25. Steer, S.J. *et al.* Single-particle behavior at N=126: isomeric decays in $^{204}$Pt. *Phys. Rev. C* **78**, 061302 (2008).

26. Podolyak, Zs. *et al*., Weakly deformed oblate structures in $^{198}$Os. *Phys. Rev. C* **79**, 031305 (2009).

27. Alkhomashi, N. *et al.* β⁻-delayed spectroscopy of neutron-rich tantalum nuclei: shape evolution in neutron-rich tungsten isotopes. *Phys. Rev. C* **80**, 064308 (2009).

28. Kurtukian-Nieto, T. *et al.* β-decay half lives of neutron-rich isotopes of Re, Os and Ir approaching the r-process path at N=126. Submitted to *Phys. Lett. B* (2010).

29. Seelmann-Eggebert, W. & Pfennig, G. Chart of the nuclides, 1st edition (1958).

**Acknowledgements.** This work was supported by grants from the Spanish Ministry of Science and Innovation, the consolider CPAN project and Conselleria de Innovación e Industria da Xunta de Galicia.

**Author Contributions.** J.B. designed and proposed the experiment. J.B. E.C., T.E. J.G., J.P., F.R. and K.H.S. mounted the experimental setup. All authors participated in the experiment. H.A., J.B. and T.K.N. did the data analysis. J.B. wrote the article with H.A., T.K.N. and K.H.S.

**Author Information.** Correspondence should be addressed to J.B. (e-mail: j.benlliure@usc.es).

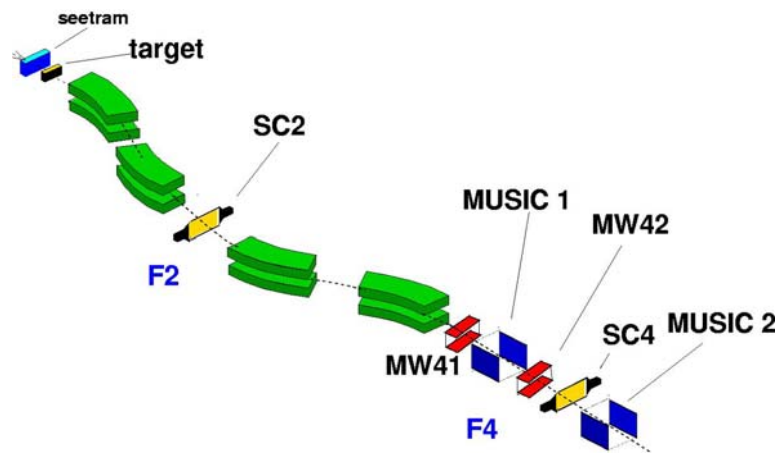

**Figure 1** Schematic representation of the experimental setup. The $^{238}$U and $^{208}$Pb projectiles accelerated at 1 A GeV at the SIS synchrotron are extracted and directed to the production beryllium target located at the entrance of the Fragment Separator. Due to the inverse kinematics, the projectile fragments are emitted forward and traverse the spectrometer.

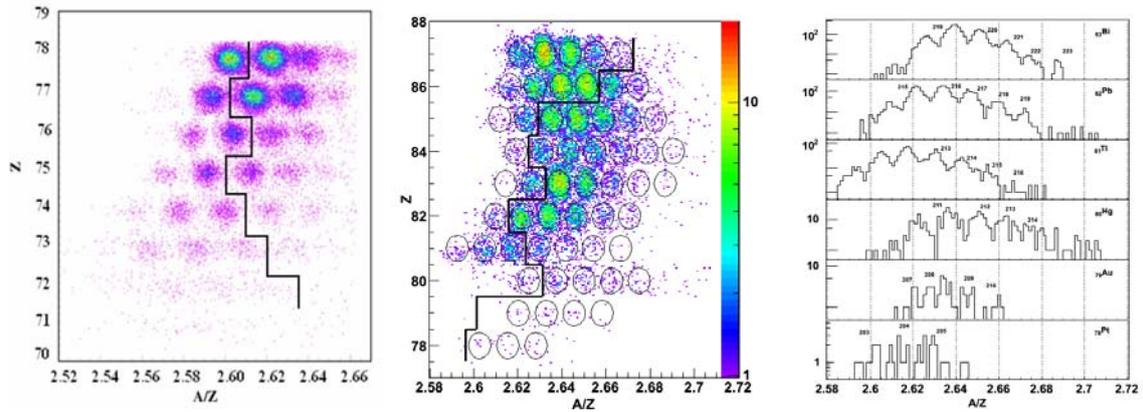

**Figure 2** Identification matrices of the nuclei produced with $^{208}$Pb (left panel) and $^{238}$U (central panel) projectiles. The solid line in these two panels indicate the limit of the most neutron-rich known nuclei before these experiments. The right panel corresponds to the A/Z distribution of the events recorded for some selected elements. The new nuclei in this range of elements are indicated by their mass number.

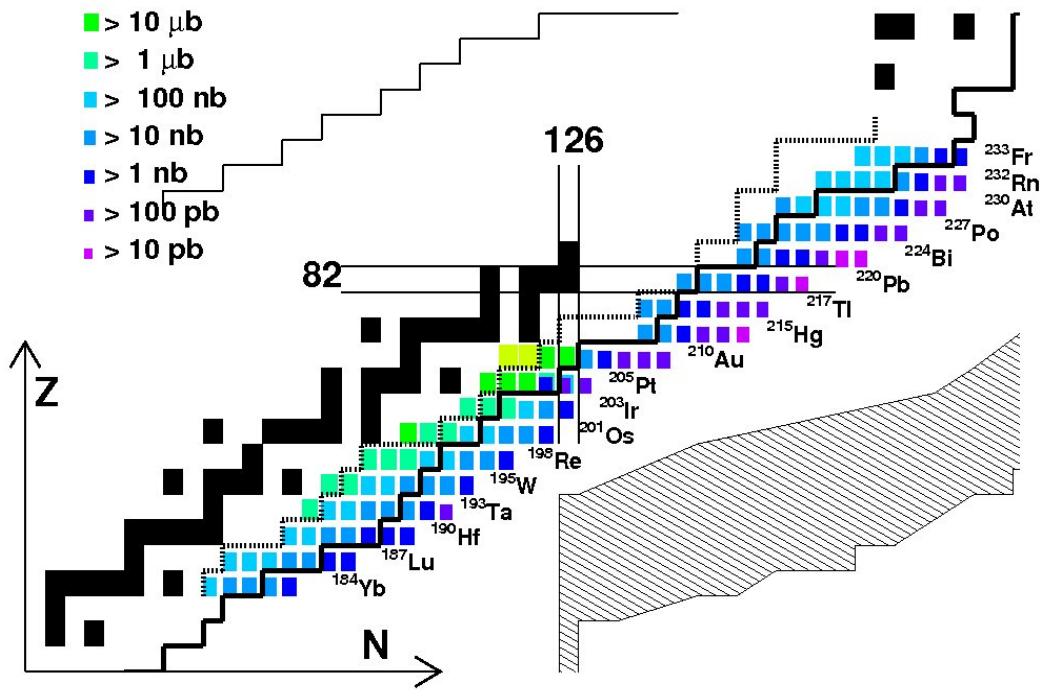

**Figure 3** Nuclei produced in this work on top of a chart of nuclides. The dotted line represents the limit of known nuclei as in 1958[29] and the thick-solid line the limit of known nuclei before these experiments.